# Automated selection of LEDs by luminance and chromaticity coordinate

Ulrich H.P. Fischer[1], Jens-Uwe Just[1], and Christian Reinboth[1]
*[1] HarzOptics GmbH*
*Dornbergsweg 2, 38855 Wernigerode, Germany*
*creinboth@harzoptics.de*

*Abstract -* **The increased use of LEDs for lighting purposes has led to the development of numerous applications requireing a pre-selection of LEDs by their luminance and / or their chromaticity coordinate. This paper demonstrates how a manual pre-selection process can be realized using a relatively simple configuration. Since a manual selection service can only be commercially viable als long as only small quantities of LEDs need to be sorted, an automated solution suggests itself. This paper intruduces such a solution, which has been developed by HarzOptics in close cooperation with Rundfunk Gernrode. The paper also discusses current challenges in measurement technology as well as market trends.**

## I. Motivation

Due to their enormous benefits, (high energy efficiency, easy handling, long service life, high insect friendlieness etc.) light-emitting diodes (LEDs) are currently seen as the most promising illuminant – the potential energy savings alone more than justify the increased use of LED. As Kim and Schubert [1] have demonstrated, a step-by-step worldwide replacement of traditional illuminants with LEDs over the next ten years wold lead to a global reduction of $CO_2$ output of about 10,6 gigatons as well as to financial savings of more than a trillion US dollars.

 LEDs play an ever-increasing role in the lighting industry. It is noteworthy, that the quality requirements for LEDs are quite different when comparing direct (street lighting, room lighting) and indirect (outline lighting) lighting applications.

When LEDs are used in direct lighting applications, e.g. a street light, the street light itself can bee seen as a point light source, meaning that the light seen by the observer ist actually a mixture of the light emitted by a multitude of LEDs within the street light [2]. Due to this, slight deviations between these LEDs concerning light intensity and / or their chromaticity coordinate would not affect the overall visual impression of the street light. Additionally, no observer with a normal eye sight could probably look directly into a street light or any other light source used for direct illumination over a prolonged period of time, making it almost impossible for such deviations to be identified by using the naked human eye. It is therefore understandable that said deviations are not an item of interest when it comes to designing direct lighting applications.

This is not the case for indirect lighting applications. When LEDs are used for the lighting of handrails in ships and airplanes or for illuminating the outline of stairs, even slight deviations in their light intensity or their chromaticity coordinate can be noticable for the human observer and therefore have a negative impact the overall visual appearance.

This is especially the case when LED light is not used just for illumination but also to create an advertising effect, e.g. indirect lighting in the 'airline's own colors' as defined in the corporate identity guidelines. Lighting scenarios like these constitute a need for the lighting equipment manufacturer to guarantee that any deviations in light intensity or chromaticity coordinate are not large enough to affect the visual appearance of the illumination, thus endangering the advertising effect.

Due to these circumstances, a small but ever-increasing demand for LED selection services has been created. Although most LED manufacturers are already offering a selection (a so-called binning) by luminance and chromaticity coordinate, these selections are quite often too rough for indirect lighting applications. The lighting industry therefore demands additional and more advanced selection services.

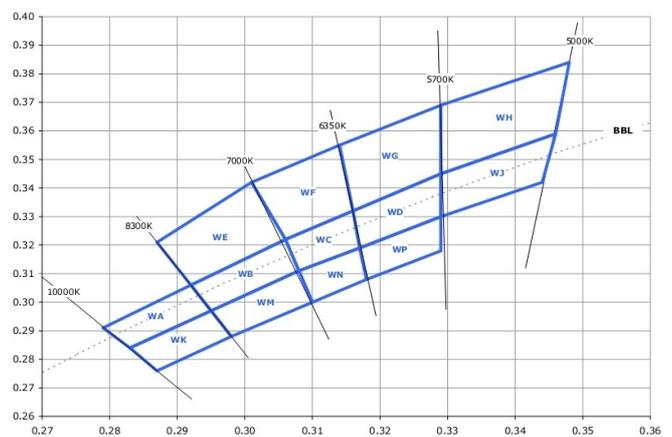

Fig. 1 Rough chromaticity coordinate screen (source: [3])

The use of a rough screen for the chromaticity data as depicted in fig. 1 has the clear disadvantage, that a human observer would still be able to visually identify deviations between LEDs who are sorted into the same screen category. In order to prevent that from happening, a more accurate screen – as shown in fig. 2 – is obviously needed.

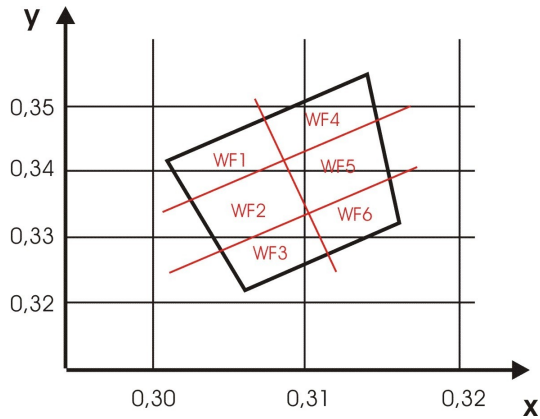

Fig. 2 More accurate chromaticity coordinate screen

## II. CHROMATICITY COORDINATE

The term 'chromaticity coordinate' refers to a coordinate in the 1931 color space as defined by the CIE – the International Lighting Comission (Commission Internationale de l'Eclarige), whose primary purpose is the development as well as the advancement of international lighting standards.

The 1931 color space shown in fig. 3 includes the entire spectrum of visible light (the visible range of the electromagnetic spectrum). Within this system, every color can be described by three corrdinates (x, y and z). Since the sum of all three coordinates is always one, only two of them are needed to identify a color. The chromaticity coordinate can therefore be defined as the exact color of the light emitted by a LED (or any other illuminant) described by x, y and z.

The color space depicted in fig. 3 is based on the 2° visual angle of the so-called standard observer. In 1964 the CIE released a second color space based on a visual angle of 10°. In both systems, the coordinates x, y and z are often called the tristimulus.

Although the tristimulus system may allow a highly precise identification of colors, it still has one major disadvantage – it does not give a realistic impression of how humans actually percept colors. As MacAdam [4] was able to prove in 1942, a standard observer will be able to detect color differences in one area of the color space while being unable to see any color differences in other areas of exactly the same size.

Fig. 4 shows the so-called MacAdam ellipses, elliptical areas within the color space in which a standard observer is unable to identify different colors.

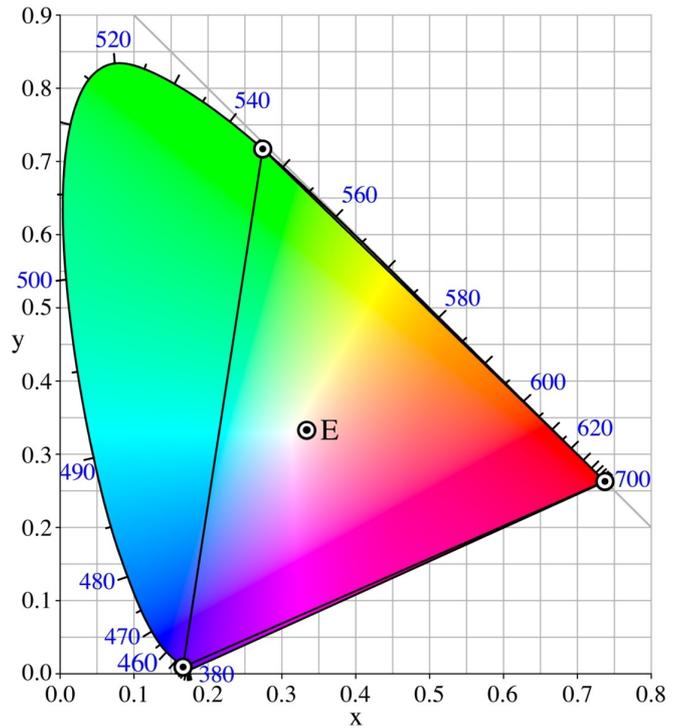

Fig. 3 CIE 1931 color space

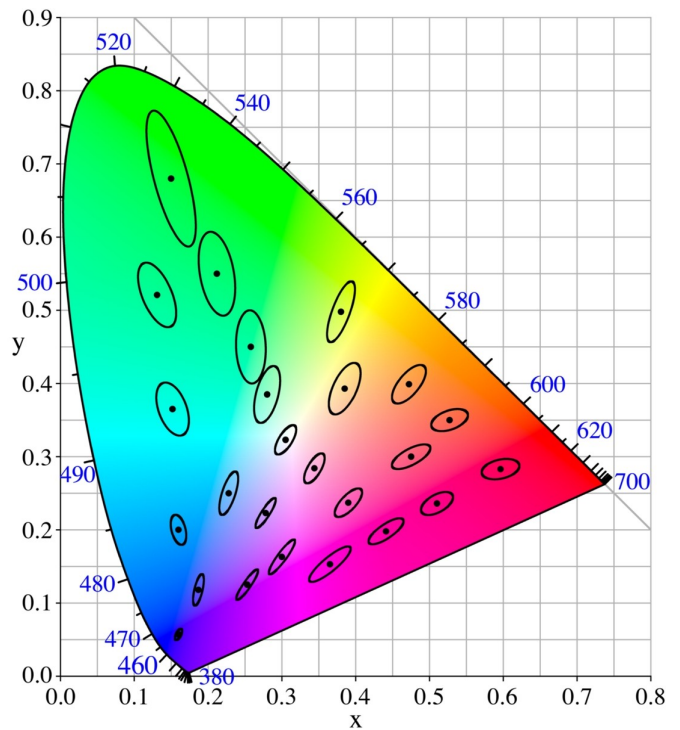

Fig. 4 MacAdam ellipses (graphic created by Torge Anders)

As can be seen in fig. 4, the MacAdam ellipses within the green or the blue range of the visible spectrum are much bigger than the MacAdam ellipses within the blue or the violet color range. This means, that a difference in chromaticity coordinates between two green LEDs is much more likely to go unnoticed than a deviation of the same extent between two blue LEDs. The demand for additional selections of blue or violet LEDs is therefore much higher than the demand for selections of green or yellow LEDs.

### III. MANUAL SELECTION

HarzOptics has been offering the selection of LEDs by luminance and chromaticity coordinate since its foundation in 2006. During the first two years, a relatively simple manual procedure was used (see overview in fig. 5).

In this configuration, the LED is manually picked up with vacuum tweezers and placed into the chuck, which is then covered by an intransparent shielding to prevent any external light from interfering with the measurement. Once the cover is closed, the LED is supplied with power and the measurement process starts. The spectrometer data (type Ando AQ 8315 with an accuracy of +/- 0,5nm) is then used to calculate the selection window using a self-developed LabVIEW software.

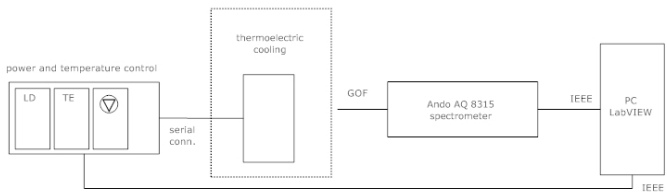

Fig. 5 Overview of the manual measurement process

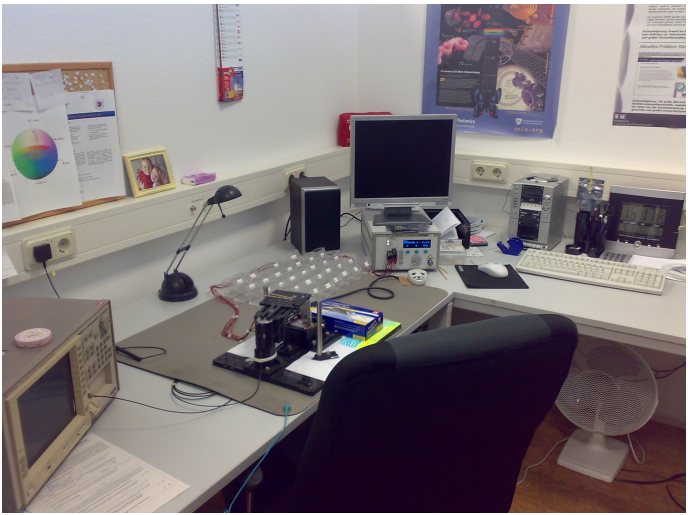

Fig. 6 Work station for manual measurements

A skilled worker can manually measure and sort between 100 and 120 LEDs per hour. A manual measurement process is therefore only expedient, if less than 150.000 LEDs are to be measured every year. Since the demand for such measurement services has increased every year and – in 2007 – had by large exceeded the number of 150.000 by large, it became necessary to develop an automated solution. This was done in 2008 by the automation experts from Rundfunk Gernrode GmbH in close cooperation with HarzOptics. The automated solution developed does not only allow the selection of more LEDs per timeframe, but has also lead to an increase in measurment and sorting precision.

### IV. AUTOMATED SELECTION

The automated selection process is about four times faster and more precise than the process of manual selection. First, the LED belt is placed into a belt feeder on the side of the machine. Inside, a grappler picks up the single LEDs one by one and places them into the chuck right in front of the integrating sphere. After the LED has been placed, it is automatically supplied with power and the measurement process is started. Since the CAS 140 CT array spectrometer is able to directly measure the chromaticity coordinate, no post measurement calculation is necessary. The LED is then picked up by the grappler and put into the corresponding compartment of a selection carousel operated via a Siemens S7 system.

The entire process takes about nine seconds, which means that about 400 LEDs can be measured in one hour with very high precision (chromaticity coordinate +/- 0,0002). The saving of time is mostly due to the faster and more precise handling of the LEDs, another important contributing factor is the high measurement speed of the spectrometer (50ms). The sytem thus allows the highly precise, low-cost selection of large amounts of LEDs (up to 3 million per year).

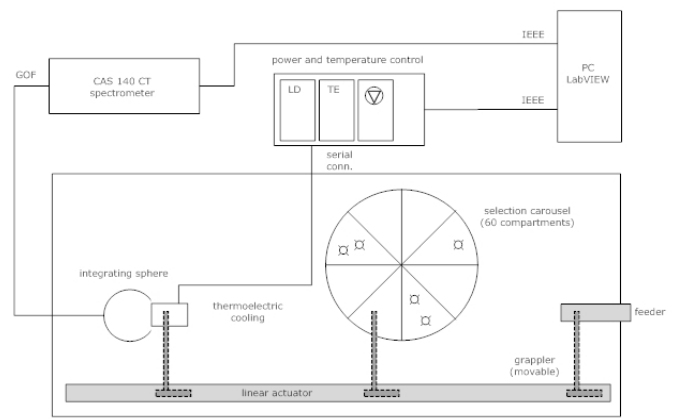

Fig. 7 Overview of the automatic measurement process

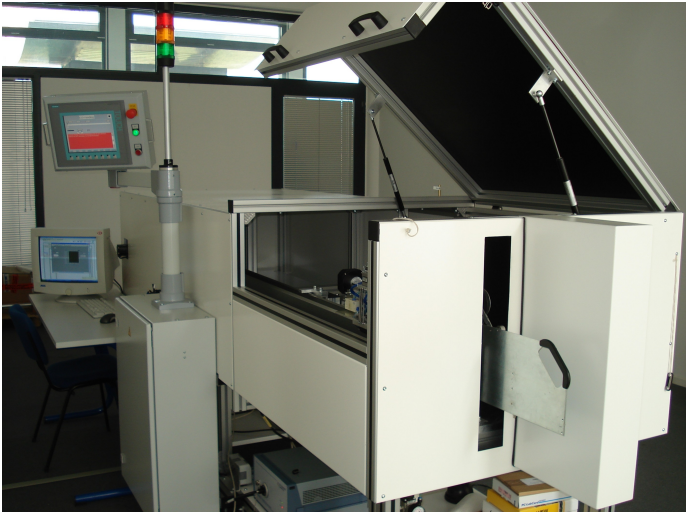

Fig. 8 Outside view of the selection apparatus

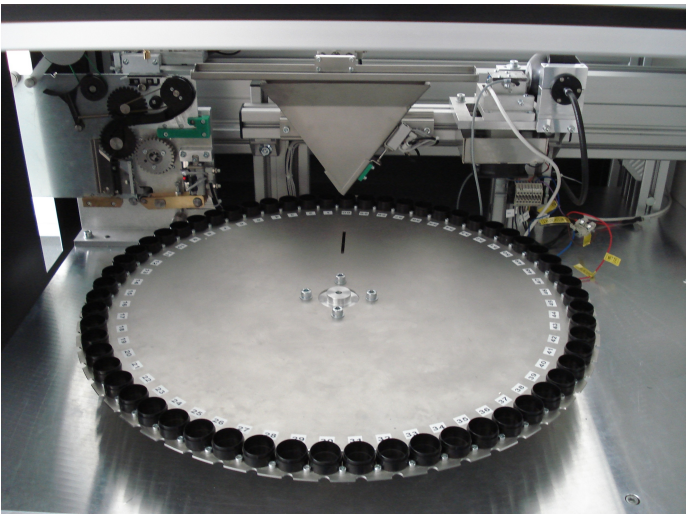

Fig. 9 Inside view of the selection apparatus

### V. LabVIEW software

In both the automated and the manual selection process the software tool LabVIEW is used to access and control the spectrometer. However, the measurement process itself differs between the automated and the manual selection.

During the manual selection process, the LED spectrum is first measured with the Ando spectrometer, the chromaticity coordinate is then calculated via LabVIEW. Since the CAS 140 CT spectrometer used for the automated selection is able to directly measure the chromaticity coordinate, no additional calculation is necessary. Aside from running LabVIEW, the computer is also needed to control the spectrometer, the power source and the temperature control. The handling of the LEDs is done by the Siemens S7 control of the selection apparatus.

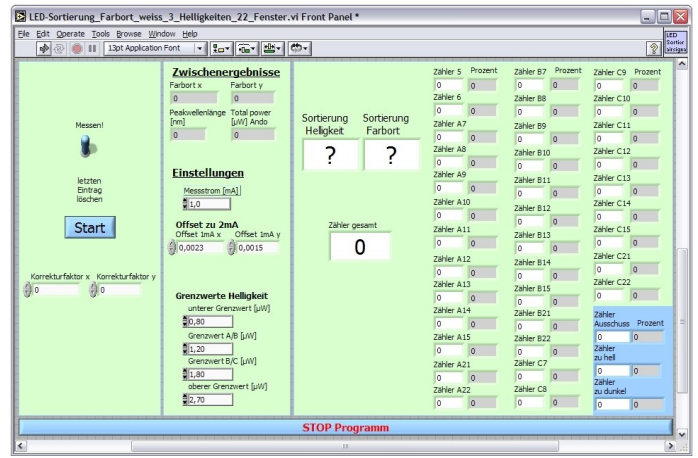

Fig. 10 LabVIEW front panel – manual selection

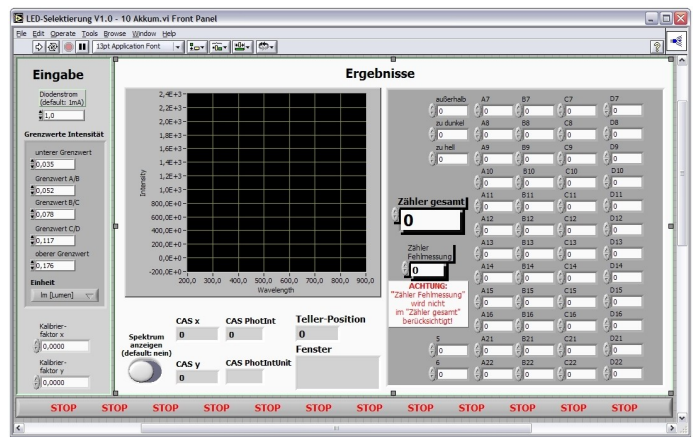

Fig. 11 LabVIEW front panel – automated selection

### VI. Conclusion and outlook

The growing interest in LED technology has already led to an increased demand for LED selection services, which can be expected to rise even further over the next two to three years. Although a manual selection is possible, it is to be regarded as highly uneconomic when large numbers of LEDs need to be processed. To solve this problem, an automated LED selection apparatus with a capacity for more than 3 million single LED measurements per year has been developed by HarzOptics and Rundfunk Gernrode. This apparatus allows a low-cost, highly precise selection of LEDs by luminance and chromaticity coordinate.

As of now, refitting the apparatus to enable the processing of different LED types still is a time-consuming procedure – the authors of this paper are currently working on ways to counter this problem and hope to have a presentable solution before the end of the year.

ACKNOWLEDGMENT

The authors would like to thank the engineering team of Rundfunk Gernrode for their tireless efforts concerning the design and construction of the automated measurement system. The authors would also like to thank Torge Anders for making his excellent graphics available under the free documentation license GNU.